# Electrochemical Synthesis of CdSe Quantum Dot Array on Graphene Basal Plane using Mesoporous Silica Thin Film Templates

Yong-Tae Kim, Jung Hee Han, Byung Hee Hong*, and Young-Uk Kwon*

*Department of Chemistry, BK-21 School of Chemical Materials Science, SKKU Advanced Institute of Nanotechnology, Sungkyunkwan University, Suwon, 440-476, Korea.*

 *Email: byunghee@skku.edu or ywkwon@skku.edu

Since the discovery of microscale single-layer graphene in 2004,[1] graphene and related materials [2] have received intensive attention as promising materials for nanoelectronics due to their fascinating electrical,[3] mechanical,[4] and chemical properties.[5,6] In addition, the recent large-scale synthesis of high-quality graphene films suggests their applications to bendable and/or stretchable transparent electrodes [7] for solar cells, sensors and displays. Surface grafting on the graphene with functional materials will be an indispensable technique towards these directions of progress. This, however, requires the problems arising from the lack of reactivity of the graphene basal plane to be resolved. A perfect graphene does not have dangling bonds on the basal plane for chemical bond formation. The chemical potential of the basal plane is lower than the edges or defects. Therefore, deposition directly on graphene results in grafting on the edges and defects only.[8] In case of using vacuum deposition techniques, this problem has been solved by modifying the graphene surface i.e., ozone treatment, forming metal underlayer, and attaching organic molecules with functional groups.[9] On the contrary, there has been no report on grafting on the basal plane by electrochemistry. Sundaram et al. reported that varying the potential during the electrochemical deposition of Pd influenced the number density of the Pd nanoparticles,[10] but the site of deposition was restricted to the edges only. None of the surface modification methods used in the vacuum deposition seems to be appropriate for the electrochemical deposition because they cannot change the potential difference between the basal plane and the edges/defects.

Here, we demonstrate that the use of a nanoporous mask can be a viable means to form a uniform nanostructured film on the graphene basal plane. We applied a mesoporous silica thin film whose pore structure is composed of about 8 nm sized vertical channels in a hexagonal symmetry (SKU-1 in ref. 11) on the graphene surface as a nanoporous mask. The nanochannels exert resistance against the diffusion of electrolytes and, thus, function as a potential-equalizer to suppress the preference for the edge and defect sites. By depositing CdSe, we formed CdSe quantum dots into a hexagonal array structure. The synthesis procedure of the present study is outlined in **Fig. 1**.

The graphene electrodes were synthesized by chemical vapor deposition of methane on thin Ni layers formed on $SiO_2$/Si substrates. [7] The Raman spectra show that the graphene films are composed of individual graphene sheets of 1~5 layers (Supporting Information). A mesoporous silica film was formed on top of this graphene/Ni bi-layer by spin-coating a precursor solution composed of tetraethoxyorthosilicate (TEOS) and a triblock non-ionic surfactant F-127 ($EO_{106}PO_{70}EO_{106}$, EO = ethylene oxide, PO = propylene oxide), followed by aging at 80°C and calcination at 400°C.[11] The pore morphology of so-synthesized mesoporous silica film, characterized by scanning electron microscopy (SEM), showed arrayed pores on the film surface, a feature expected for the SKU-1 film (Supporting Information). By using the SKU-1-coated graphene as a working electrode, CdSe was deposited under a constant potential of -0.7 V at 50°C. After the deposition, the SKU-1 template was removed by dissolving in an aqueous HF solution. We also prepared a CdSe film grown on a graphene/Ni bi-layer without the SKU-1 coating following the same procedure except for the formation of SKU-1. The energy dispersive X-ray spectroscopy (EDS) data on these samples show the presence of Cd and Se (Supporting information).

**Fig. 2** shows the morphologies of the two CdSe films. The atomic force microscopy (AFM) image of the film without the SKU-1 template (**Fig. 2a**) shows that CdSe was deposited into lines forming closed loops enclosing domains of various areas from 0.3 x 0.3 to 1.5 x 1.5 μm$^2$.

Based on the literature data on graphenes, these loops are probably the edges around the multiple layered graphene sheets.[7] The height of the CdSe deposit varies from 7 to 9 nm by the AFM height scan. The SEM images on this sample with a higher magnification (**Fig. 2b**) reveals that the loops are composed of small CdSe particles and that there are CdSe deposits on the basal plane also, probably at the defect sites. Compared with the literature data on the electrochemical deposition of Pd on graphene,[10] our data show a much higher density. Probably, the Ni underlayer provides conduction paths circumventing the problem of the high resistance of the graphene basal plane.

The images of the SKU-1-templated CdSe film show that the template brings about quite a dramatic effect. The wide area AFM image (**Fig. 2c**) shows that the loop-like structure seen in **Fig. 2a** on the CdSe film formed without SKU-1 template has disappeared. Instead, the whole surface of graphene is covered with a very fine structured film. The height variation plot across the image shows that the film is composed of even sized particles of about 5-12 nm in height. The SEM image (**Fig. 2d**) shows that there are arrays of dots into a hexagonal symmetry and the dot-to-dot distance is measured to be 14 nm. These morphological characteristics agree well with the pore structure of the SKU-1 template.

**Fig. 3** shows the transmission electron microscopy (TEM) images of the CdSe quantum dot array grown with the SKU-1 template. The TEM sample was prepared by removing the Ni underlayer by treating with a $FeCl_3$ solution.[7] These images clearly show that this film is composed of nanoparticles in a hexagonal array structure. The dot size was measured to be 10 nm and the dot-to-dot separation 14 nm, confirming the above observations with AFM and SEM. The selected area electron diffraction pattern of this sample (**Fig. 3c**) shows a ring pattern with d = 0.26, 0.22, and 0.18 nm that match well with the (102), (110), and (112) peaks of the hexagonal form of CdSe. The magnified image in **Fig. 3b** also shows lattice fringes that can be explained with the CdSe lattice.

Our data clearly show that (i) the use of graphene with a Ni underlayer and (ii) the use of SKU-1 as a nanoporous mask, combined, make it possible to synthesize CdSe quantum dots into an array structure on the graphene basal plane. We can explain the effects of these two measures as the decrease of the difference of the resistances between the edge sites and the basal plane. The Ni underlayer provides conduction paths so that it can reduce the resistance to the points on the basal plane. The nanoporous SKU-1 mask functions as a diffusion barrier for the incoming ions in the electrolyte solution to the extent to override the conductivity difference between the edges/defects and the basal plane. Furthermore, the regular pore structure of SKU-1 functions as a template to deposit quantum dots into an array structure. The Raman spectroscopy data in **Fig. 4** show that the CdSe quantum dots induce red-shifts of the G and 2D bands of graphene by 5 and 9 cm$^{-1}$, respectively. In order to verify that these changes are indeed induced by the CdSe quantum dots, we treated graphene films in various ways similar to the method used to form CdSe quantum dots but without the deposition of CdSe and measured their Raman spectra (Supporting Information). None of such graphene films produced the similar changes both in direction and magnitude in the Raman peaks, which leads us to conclude that the large red-shifts are due to the CdSe quantum dots. The G and 2D band positions are known to be sensitive to the electron or hole doping.[12, 13] The observed red-shifts imply that the graphene sheet is doped with electrons. From the magnitude of the G-band shift, the density of charge carrier (electron) is estimated to be $3.0 \times 10^{12}$ cm$^{-2}$ which corresponds to an up-shift of the Fermi level by ~100 meV from that of pure graphene. This level of n-doping is larger than that reported for the viologen-modified CNT which showed a G-band shift of 2 cm$^{-1}$.[14] These data are significant in view that most of the surface modifications of graphene in the literature result in p-doping.[15, 16] Therefore, the n-doping characteristics of the CdSe quantum dot-modified graphene may find applications to engineer the band-gap structures of graphene nanostructures for various electronic applications.

The mechanism of the n-doping by CdSe quantum dots is quite intriguing. Because the work function of graphite, reported to be 4.6-5.0 eV,[17, 18] is located at the HOMO-LUMO gap of CdSe,[19] it seems possible that the electrons of the conduction band of CdSe quantum dots flow into graphene. Very recently, Farrow and Kamat reported that CdSe quantum dots on CNTs can inject electron to CNTs by the same mechanism.[20] On the other hand, Zhou et al. reported that an epitaxially grown graphene on a SiC substrate was n-doped and discussed that this was probably associated with the surface charges at the interface between graphene and SiC substrate,[21] which mechanism might be responsible for our system. Alternatively, although bulk CdSe is known to be stoichiometric with Cd:Se = 1:1, it is possible to make CdSe nanoparticles either Cd- or Se excess.[22] Therefore, if the CdSe quantum dots in our samples are Se excess and, thus, are n-doped, they may be able to cause the n-doping of the graphene. Further studies are required to better understand the effect of the CdSe quantum dots to the red-shifts of the Raman peaks.

In summary, we demonstrate, for the first time, a method that can produce a homogenous nanostructured thin film with an ordered nanostructure on the basal plane of a graphene by electrochemistry. This method enables the direct contacts between CdSe quantum dots and graphene films without intervening layers, possibly leading to the enhancement of electrical and mechanical properties.[23] With these potentials, the present method is expected to open a new horizon of exploration on graphene in many fields.

*Experimental*

*Synthesis of graphene*: The graphene/Ni bi-layer was synthesized by the chemical vapor deposition method reported in Ref. 7. A thin layer of nickel of thickness less than 300 nm was formed on a $SiO_2$/Si substrate using an electron-beam evaporator and was placed in a quartz tube under an argon atmosphere. The temperature was raised to 1000°C and a reaction gas mixture ($CH_4$:$H_2$:Ar = 550:65:200 standard cubic centimetres per minute) was flown. Then the substrate was cooled to room temperature at the rate of ca. 10 °C sec$^{-1}$. A graphene film was formed on top of the nickel layer.

*Synthesis of mesoporous silica thin film*: A stock solution for SKU-1 was prepared by dissolving TEOS (99.999%) and F-127 in a mixed solution of diluted aqueous HCl and absolute ethanol to make the composition TEOS : F-127 : HCl : H$_2$O : EtOH = 1 : 6.60x10$^{-3}$ : 6.66x10$^{-3}$ : 4.62 : 22.6 (molar ratio). The solution was stirred at 20–25°C for 12–20 h under a controlled relative humidity of below 20 % before use. The stock solution was spun-coat on the graphene/Ni bi-layer substrate, aged at 80°C, and calcined at 400°C to form a nanoporous silica thin film.

*Electrochemical synthesis of CdSe quantum dots*: A corner of the SKU-1 film on the graphene/Ni bi-layer was etched out to reveal the graphene surface to make an electrical contact with a wire for the following electrochemical deposition. Using a SKU-1-coated graphene as the working electrode and a Ag/AgCl reference and a Pt counter electrodes in a three-electrode cell and an aqueous electrolyte solution (0.3 M CdSO$_4$, 0.003 M SeO$_2$), CdSe was deposited under a constant potential of -0.7 V at 50 °C.[24] After the deposition, the SKU-1 template was removed by dissolving in an aqueous 0.2 wt% HF solution. A CdSe film on a graphene/Ni bi-layer without the SKU-1 coating was also prepared following the same procedure except for the formation of SKU-1.

*Characterization*: SEM (JEOL 7100F, 5~10kV), AFM (Park systems XE-NSOM; non-contact mode), and TEM (JEOL JEM 3010, 300kV) were used to characterize the samples. Raman spectra were recorded with a Ranishaw RM 1000-Invia by using a 514 nm excitation and a notch filter of 50 cm$^{-1}$. The spectral resolution is 0.2 cm$^{-1}$.

*Acknowledgements*
This work was supported by the Korea Science and Engineering Foundation Grant funded by the Korean Government (MOEHRD) (2009–008–1018). We thank CNNC for AFM and CCRF for SEM/TEM images.

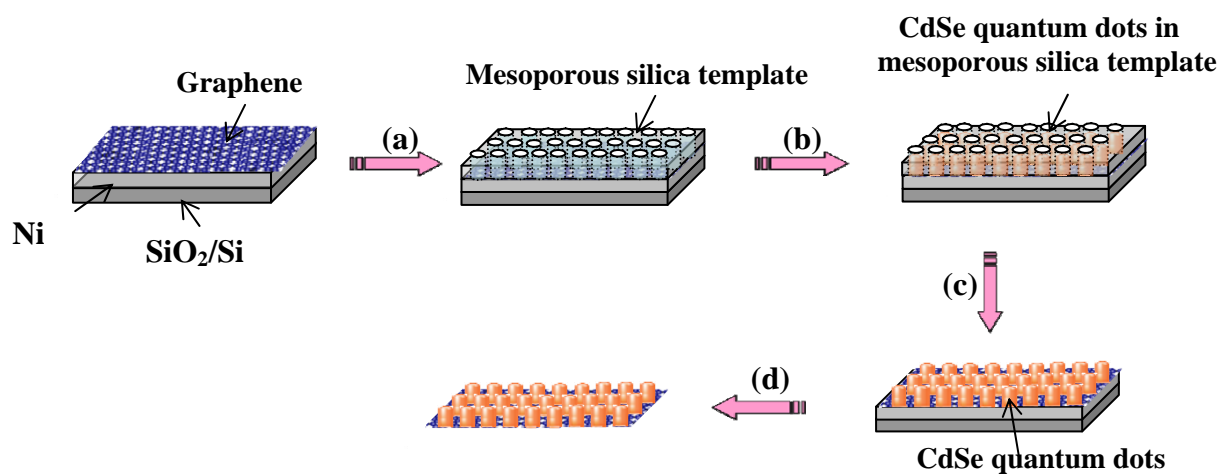

**Figure 1.** Procedure to synthesize a CdSe quantum dot array on the basal plane of a graphene sheet using a mesoporous silica thin film as a template: (a) Formation of a mesoporous silica film on the graphene surface by spin-casting, aging and calcination, (b) electrochemical deposition of CdSe onto the graphene surface through the pores of the mesoporous silica film template, (c) removal of the mesoporous silica template, and (d) removal of the Ni layer underneath the graphene sheet.

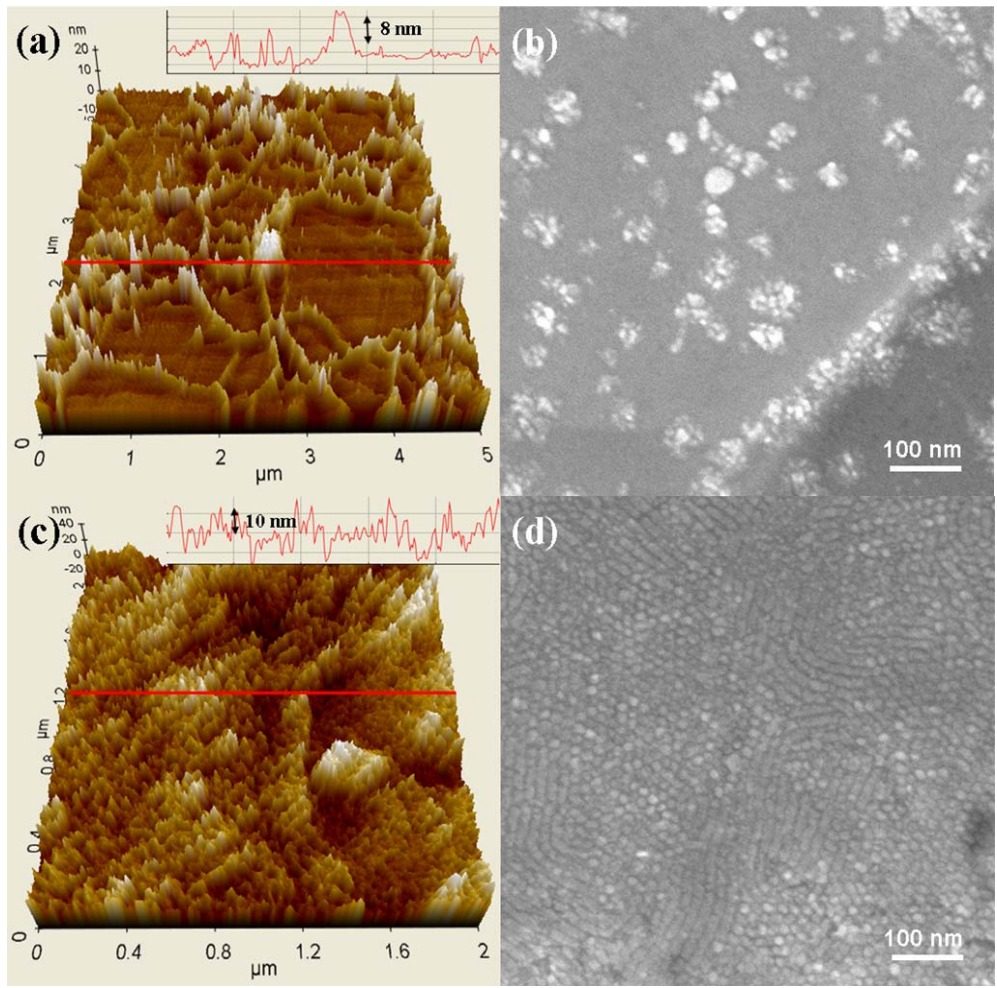

**Figure 2.** (a) AFM and (b) SEM images of CdSe grown on a pristine graphene without a mesoporous silica film template, (c) AFM and (d) SEM images of CdSe quantum dot array grown on a graphene by using a mesoporous silica film template. Insets in (a) and (b) are the height profiles taken along the red lines marked in the corresponding AFM images.

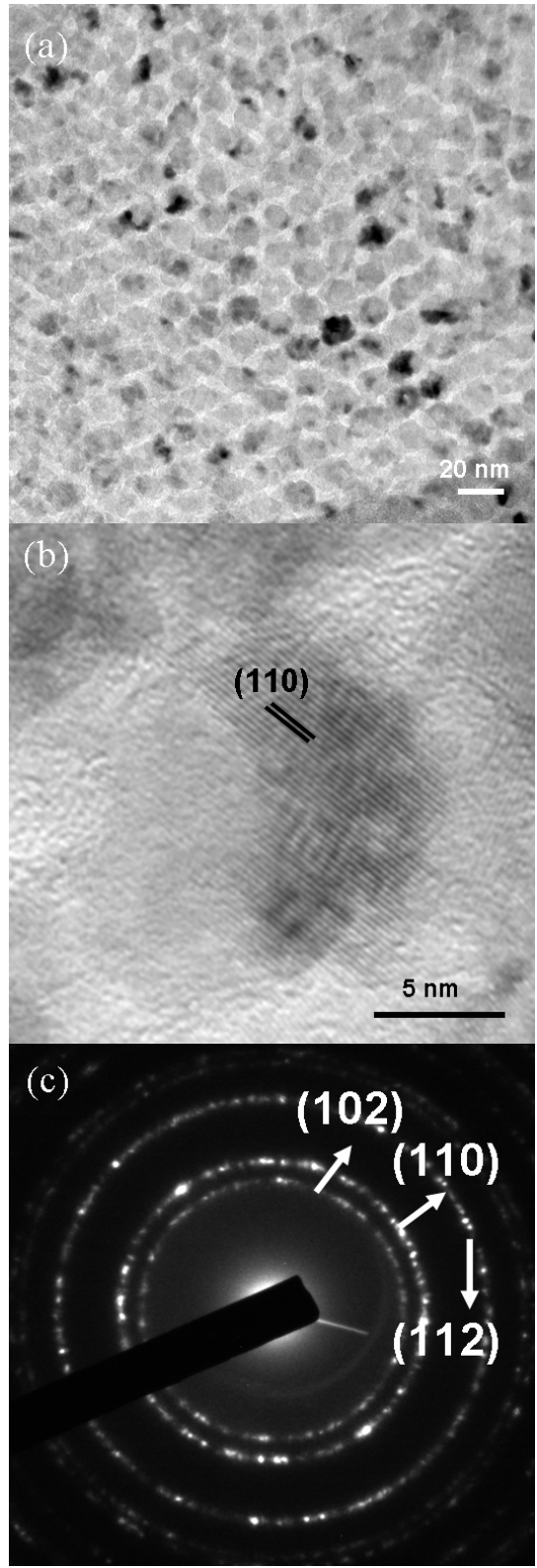

**Figure 3.** (a) High resolution TEM image of a CdSe quantum dot array, (b) enlarged view of (a) to reveal the lattice fringe, and (c) SAED pattern.

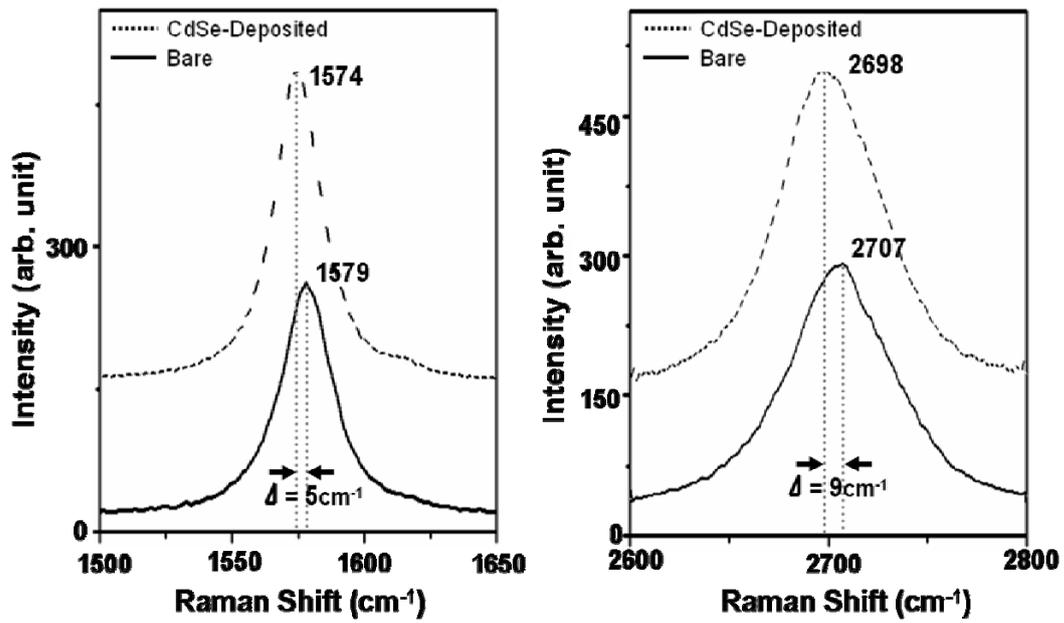

**Figure 4.** (a) G-band and (b) 2D-band Raman Spectra of a graphene with CdSe quantum dots deposited on the surface (dash lines) and a bare graphene film (solid lines).

# Supporting Information

# Electrochemical Synthesis of CdSe Quantum Dot Array on Graphene Basal Plane using Mesoporous Silica Thin Film Templates

Yong-Tae Kim[1], Jung Hee Han[2], Byung Hee Hong[1,2]*, Young-Uk Kwon[1,2]*

. [1] Department of Chemistry, BK-21 School of Chemical Materials Sciences, &

[2] SKKU Advanced Institute of Nanotechnology, Sungkyunkwan University,

Suwon 440-746, Korea.

## 1. Raman characterization of a few-layer graphene film

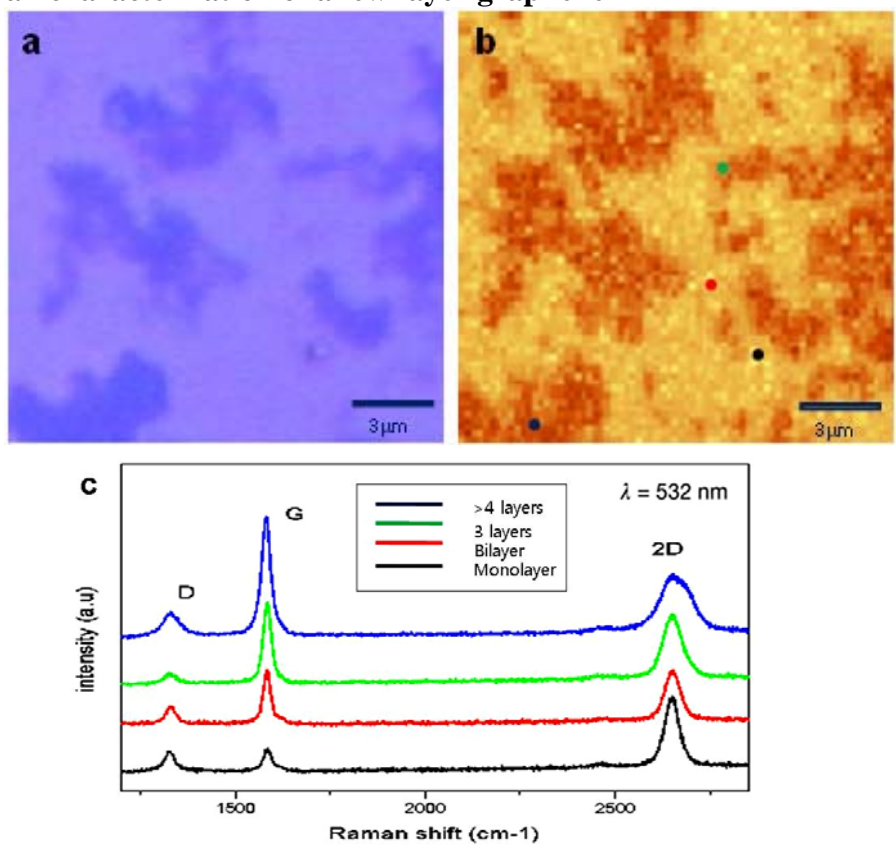

**Figure S1**. (a) Optical microscopic image of graphene films transferred on a SiO$_2$ substrate. (b) Confocal scanning Raman image corresponding to (a). (c) Raman spectra taken from the corresponding colored spots in (b).

## 2. Synthesis of a mesoporous template (SKU-1) on a graphene film

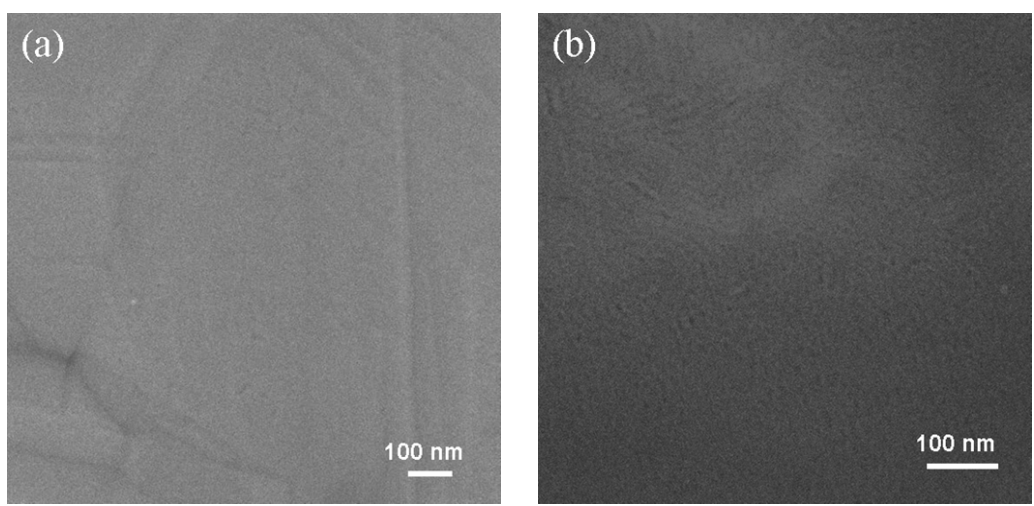

**Figure S2**. SEM images of (a) a pristine graphene sheet and (b) a mesoporous silica film formed on the surface of a graphene sheet.

## 3. EDX spectra of CdSe nanoparticles electrodeposited on a graphene film.

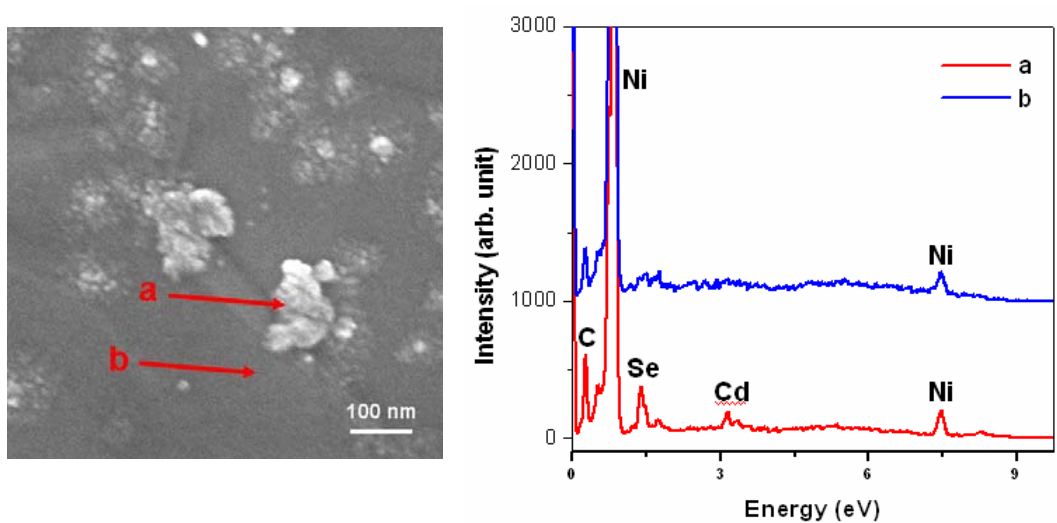

**Figure S3**. EDX spectra of graphene and CdSe nanoparticles on a CdSe-deposited graphene film without a mesoporous silica thin film template (**a, b**). The Ni peaks are due to the Ni underlayers of the grapheme films. An EDX equipped on a SEM was used.

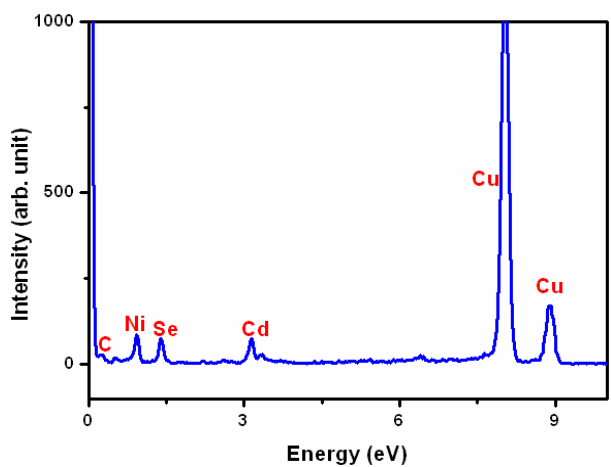

**Figure S4**. EDX spectrum of CdSe nanoparticles on a CdSe-deposited graphene film with a mesoporous silica thin film template. The Cu peaks are from the Cu-grid for the TEM measurement. An EDX equipped on a TEM was used.

**4. Raman spectra of graphene films treated in various ways.**

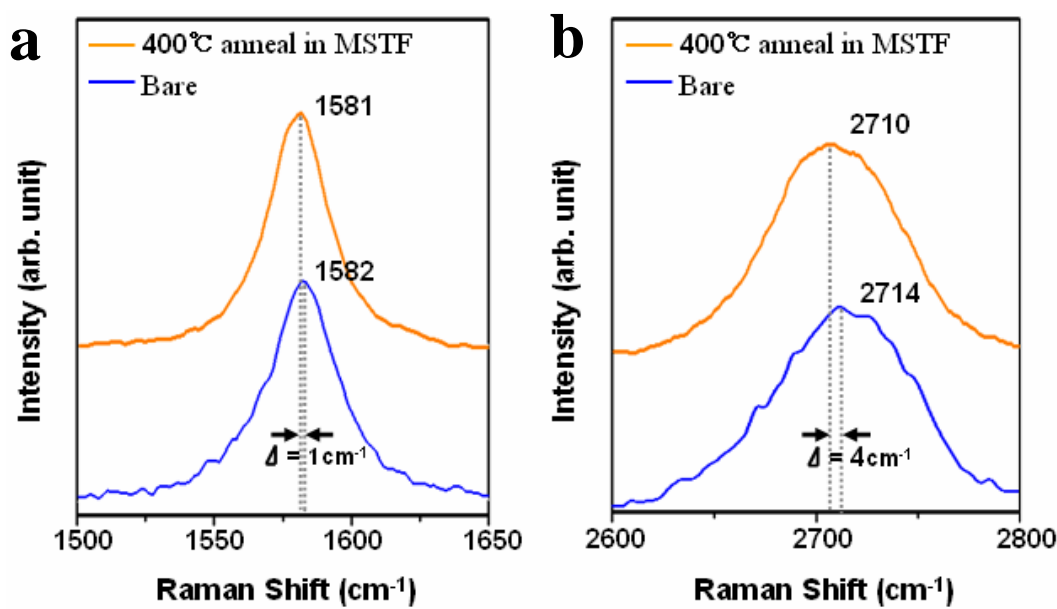

**Figure S5**. (a) G-band and (b) 2D-band Raman Spectra of a graphene on which a mesoporous silica thin film was formed and removed (treated in the same way the one in Figure 4 except the formation of CdSe quantum dots; orange lines) and a bare graphene film (blue lines).

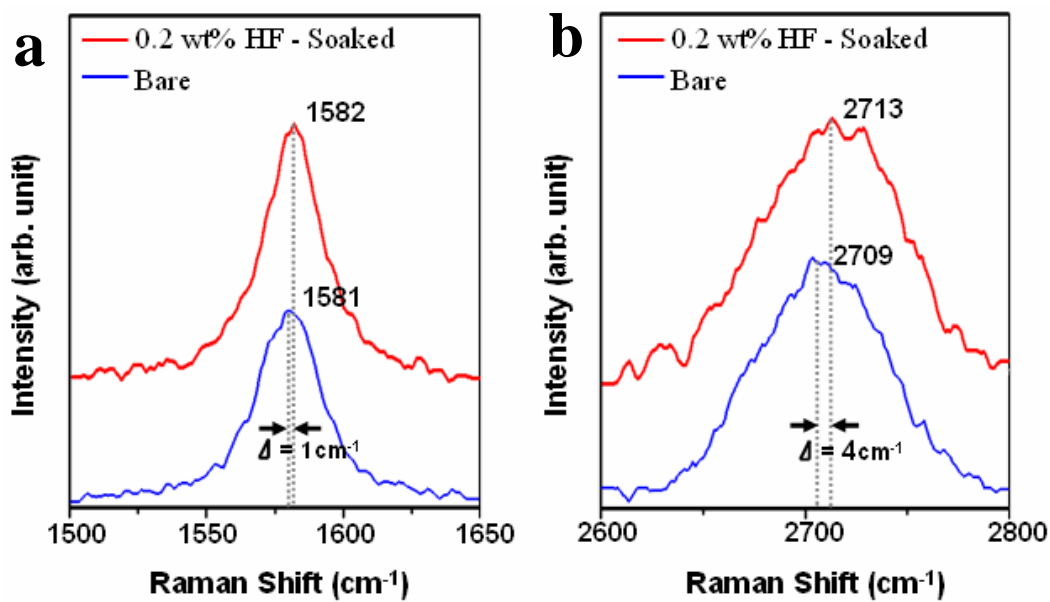

**Figure S6**. (a) G-band and (b) 2D-band Raman Spectra of a graphene before (blue lines) and after (red lines) a treatment by soaking in a 0.2 wt% HF solution. The spectra were taken on the same area.

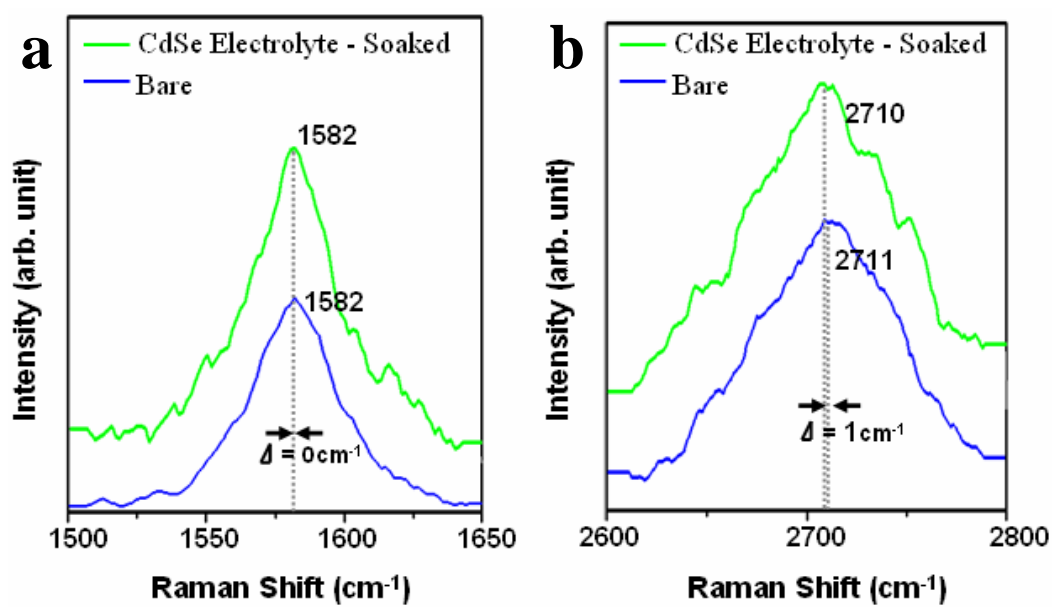

**Figure S7**. (a) G-band and (b) 2D-band Raman Spectra of a graphene before (blue lines) and after (green lines) treatment by soaking in a electrolyte solution used to deposit CdSe quantum dots. The spectra were taken on the same area.

**5. CdSe nanoparticles electrodeposited of a graphene film.**

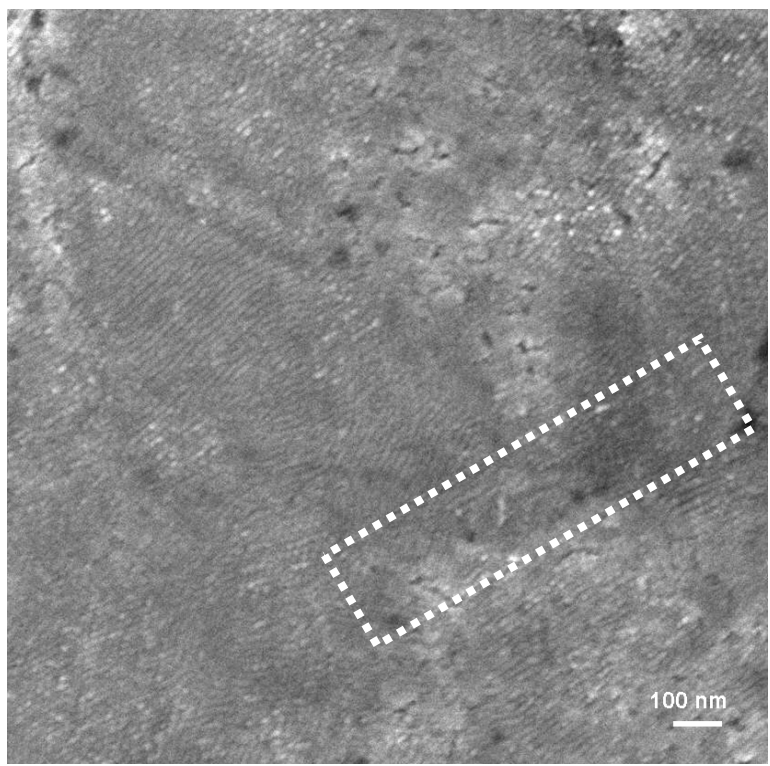

**Figure S8**. SEM image of CdSe quantum dots on a graphene film over a large area of 1.6 x 1.6 µm$^2$. The area is large enough to reveal the edges of graphene sheets. The absence any overly deposited CdSe indicates that the preferential deposition on the edge sites is suppressed by our method. The boxed area indicates a boundary between different graphene sheets. Note that the coverage by the CdSe quantum dots in this region is uniform, indicative the absence of the edge effect to the growth of CdSe quantum dots.